# Optical Characterization of Electro-spun Polymer Nanofiber based Silver Nanotubes

S. Ganganagunta[a]

[a]Department of Physics, Koneru Lakshmaiah University, Guntur, Andhra Pradesh 522502, India

**Abstract:**

Nanotubes of various kinds have been prepared in the last decade, starting from the discovery of carbon nanotubes. Recently other types of nanotubes including metallic (Au), inorganic ($TiO_2$, $HfS_2$, $V_7O_{16}$, CdSe, $MoS_2$), and polymeric (polyaniline, polyacrylonitrile) have been produced. Herein we present a novel synthetic procedure leading to a new kind of porous, high-surface-area "nanoparticle nanotubes" (NPNTs). This study characterizes the synthesized silver nanotubes at optical wavelengths. The absorption spectrum of PAN washed silver nanotubes shows an extended absorption peak at visible wavelengths ranging from 350 – 700 nm. In addition, the absorption spectrum of randomly oriented silver nanotubes showed plasmonic behavior, indicating high efficient surface enhanced Raman scattering (SERS) performance.

**Introduction:**

Nanotubes of various kinds including carbon, metallic[1-3] (e.g., Au), inorganic (e.g., $TiO_2$, $HfS_2$, $V_7O_{16}$, CdSe, $MoS_2$), and polymeric[4-6] (e.g., polyaniline, polyacrylonitrile) were synthesized by several methods; hydrothermal synthesis, surfactant-assisted synthesis, and decomposition in $H_2$ or electron radiation. A common method for producing nanotubes is template synthesis in nanoporous membranes. Template synthesis of nanotubes is achieved by using different strategies such as electrochemical deposition, electro-less deposition, polymerization, sol–gel deposition, or chemical vapor deposition (CVD) in the nanoporous templates. The "nanoparticle nanotube" (NPNT) are prepared by assembly of gold nanoparticles on the pore walls of a silane treated nanoporous alumina membrane template, accompanied by spontaneous room-temperature coalescence of the bound nanoparticles[1]. Under well-defined conditions this process results in a solid tubular structure spanning the entire pore length. Self-sustained NPNTs, which preserve the nanoparticle morphology, are obtained by template dissolution.

**Design:**

Fabrication of Solid Polyacrylonitrile (PAN) nanofibers of less than 500 nm using electrospinning method. Deposition of 100 nm thick uniform silver coating on PAN fibers using dipping process.

Washing off PAN with Dimethylformamide (DMF) to obtain desired hollow silver nanotubes of 800 nm diameter. Figure 1 shows the solid PAN nanofibers and silver coated solid PAN fibers and ultimately the hollow silver nanotubes.

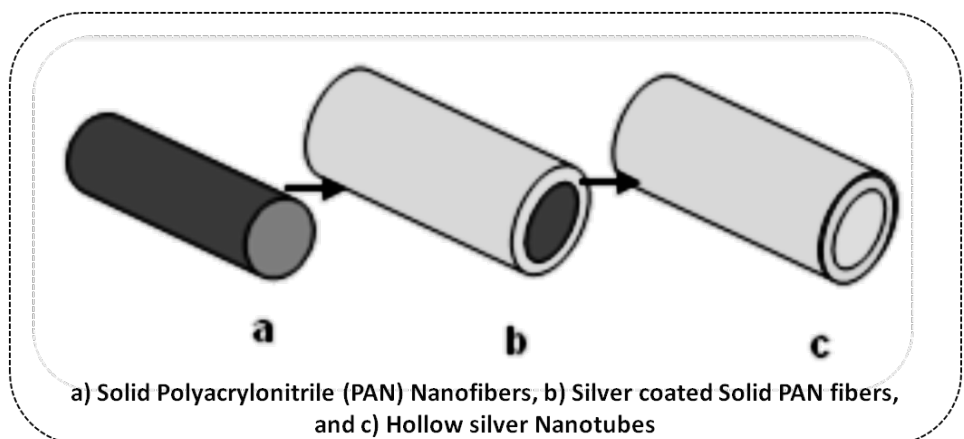

a) Solid Polyacrylonitrile (PAN) Nanofibers, b) Silver coated Solid PAN fibers, and c) Hollow silver Nanotubes

Figure 1: The schematic of silver nanotubes preparation

## Synthesis / Fabrication:

### 1. Electrospinning

A common, simple method used for producing randomly oriented polymer nanofibers is electrospinning (a drawing process based on electrostatic interactions). Figure 2a shows the schematic illustration of the basic setup for electrospinning. It consists of three major components: a high-voltage power supply, a spinneret (a metallic needle), and a collector (a grounded conductor). The spinneret is connected to a syringe in which the polymer solution is hosted. With the use of a syringe pump, the solution can be fed through the spinneret at a constant and controllable rate. When a high voltage is applied, the pendent drop of polymer solution at the nozzle of the spinneret will become highly electrified and the induced charges are evenly distributed over the surface. As a result of two major types of electrostatic forces: the electrostatic repulsion between the surface charges; and the Coulombic force exerted by the external electric field, the liquid drop will be distorted into a conical object commonly known as the Taylor cone (as shown in the inset of Figure 2a). By electrospinning process, we obtained solid polyacrylonitrile (PAN) nanofibers (of 500nm diameter) in 2-4 minutes as shown in Figure 2b. It has an ideal generally been accepted that there is coordination between silver ions and nitrogen loan pair of PAN. The coordination between cyano nitrogen and silver ion makes PAN carrier of silver ion to be deposited on PAN fibers.

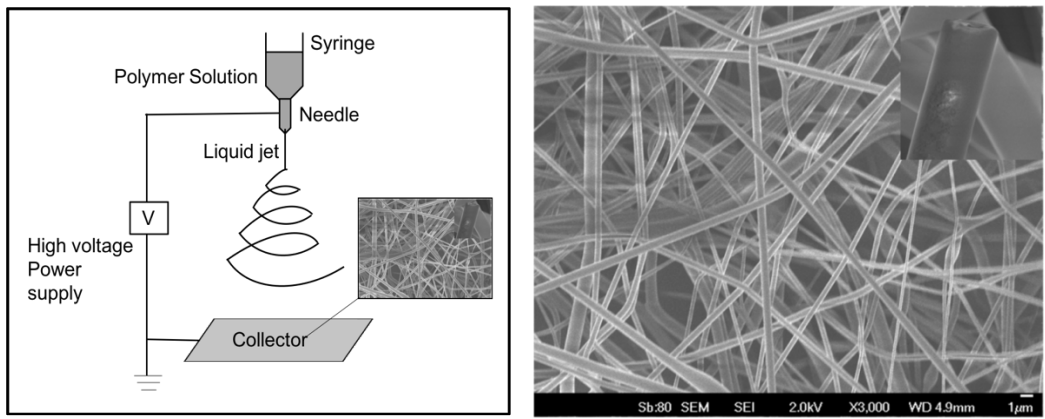

Figure 2: a) Schematic illustration of the basic setup for electrospinning, b) SEM image of Polyacrylonitrile (PAN) fibers.

## 2. Silver Deposition

Silver coating of PAN fibers can be achieved by simple electro less deposition, dipping method (as shown in the Figure 3), which will take just 3 minutes. Sensitizing solution of 100 ml. was prepared by dissolving 0.05 g of $PdCl_2$ and 2.5 g of $SnCl_2$ in 100 ml. of dilute hydrochloric acid (with a pH of 1.0). The polymer fibrous mat was rinsed with de-ionized/distilled $H_2O$, and with the sensitizing solution. The Ag plating solution was prepared by dissolving 2.4 g of $AgNO_3$ in 800 ml of pure $H_2O$. A mixture of 6% $NH_4OH$ and 0.44 g NaOH was added to the solution; in order to complex the silver and to improve the adhesion, durability, and hardness of the Ag layer[7]. The polymer fibrous mat is immersed in silver nitrate plating solution. The reducer solution was prepared by dissolving 0.08 g of sodium ethylene-diamine tetra acetic acid, ($Na_2EDTA$) and 0.56 g of dextrose in pure $H_2O$. When the reducer solution is added to the fibers a thin Ag film is deposited at the surface of the fibers. Uniform silver coating can be obtained by applying wetting solution prior to the silver deposition.

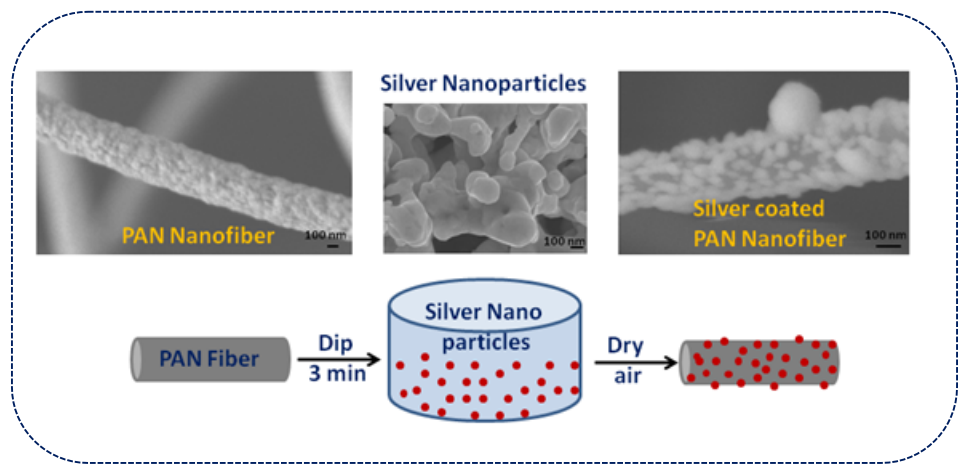

Figure 3: The schematic of silver deposition process

### 3. Polymer Dissolution

The desired hollow silver nanotubes of 100 nm can be obtained by dissolving the silver coated PAN fibers in Dimethylformamide (DMF). The polymer coating can be washed off either by sonicating the sample for 5-10 min or by just disolving the metal coted fiber mat in DMF, to obtain silver nanotubes of less than 1 µm diameter. The SEM image of PAN washed silver nanto tubes were shown in Figure 4.

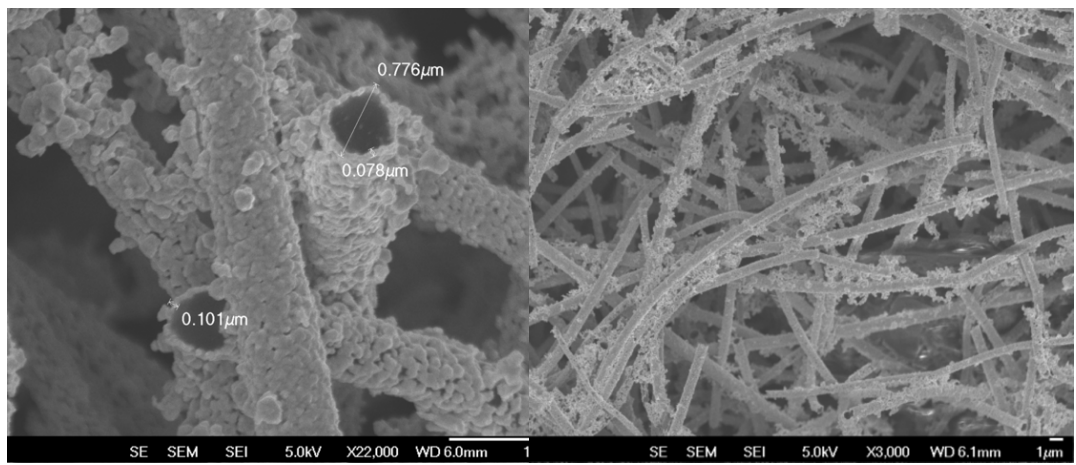

Figure 4: SEM image of PAN washed silver nanotubes using Dimethylformamide (DMF)

This is a very simple and novel technique that takes less than 10 minutes (including the preparation of solutions). The inner diameter of the Ag nano tubes can be varied by changing the diameter of the solid PAN fibers, i.e by controlling the needle tip size as shown in the electrospinning setup. Also, the thickness of the silver nanotubes can be varied by changing the silver deposition time, i.e by increasing/decreasing the dipping time. Whereas the conventional methods reported in the literature, for the fabrication of metal nanotubes are tedious and take more than 4 hours (excluding the fabrication/preparation of membrane). Due to the difficulty involved in the fabrication of smaller pore size alumina membrane (pore diameters less than 200nm), it is hard to achieve thin metal nanotubes with smaller bore diameters.

## Characterization:

The absorption spectrum of PAN washed silver nanotubes showed an extended absorption peak from 350 – 700 nm (instead of a sharp peak at 398 nm), as shown in the Figure 5. In other words, the Absorbance Vs Wavelength graph showed the plasmonic behavior of the randomly oriented silver nanotubes which is related to the surface enhanced Raman scattering (SERS). Therefore, Ag nanotubes showed an improved SERS performance. Surface enhanced Raman scattering (SERS) is known as a powerful tool for chemical and biological detection, providing molecular structural information together to make a huge impact in life sciences, environmental monitoring, and homeland security. Also, with ultrasensitive detection limits, and even single molecule sensitivity. In addition to these it has been applied to Anthrax detection, cancer detection[8-10], chemical warfare-stimulant detection, SERS performance[11], in vitro and in vivo sensing[12-14].

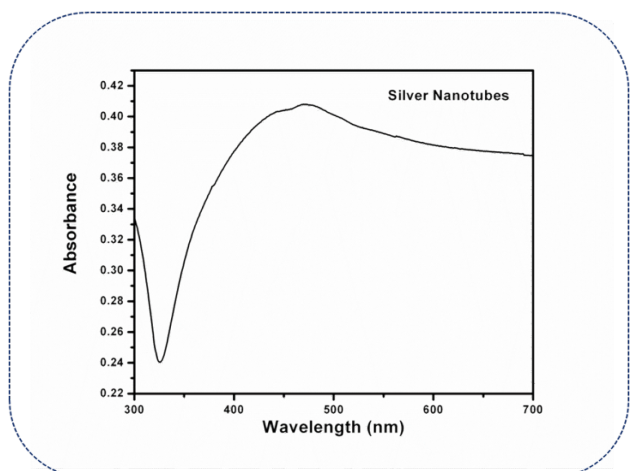

Figure 5: Absorption Spectrum of PAN washed Ag Nanotubes

## Conclusion:

In conclusion, a novel procedure is presented, providing metal "nanoparticle nanotubes" (NPNTs) that combine nanotube geometry with nanoparticle morphology and properties (e.g., high surface-to-volume ratio; surface plasmon optical absorption). The process involves dipping the PAN nanofibrous mat in a metal (Ag) colloid solution, followed by spontaneous room-temperature coalescence of the surface-confined nanoparticles to yield uniform metal coating on polymer fibers. The metal nanotubes can be obtained as free-standing NPNTs by dissolution of the polymer fibers. The mechanism of NPNT formation is not yet fully understood and is currently beingstudied. The new process opens the way to the synthesis of novel kinds of nanomaterials that have a tube geometry, high surface area, mechanical stability, electrical conductivity, and distinct optical properties. The synthesis of various other metal NPNTs, as well as morecomplicated structures such as bimetallic NPNTs, is being investigated.

## References:

1. Lahav, M., Sehayek, T., Vaskevich, A., & Rubinstein, I. (2003). Nanoparticle nanotubes. Angewandte Chemie International Edition, 42(45), 5576-5579.
2. Doradla, P., Joseph, C. S., Kumar, J., & Giles, R. H. (2012). Characterization of bending loss in hollow flexible terahertz waveguides. Optics express, 20(17), 19176-19184.
3. Doradla, P., Joseph, C. S., Kumar, J., & Giles, R. H. (2012, February). Propagation loss optimization in metal/dielectric coated hollow flexible terahertz waveguides. In SPIE OPTO (pp. 82610P-82610P). International Society for Optics and Photonics.
4. Li, D., & Xia, Y. (2004). Electrospinning of nanofibers: reinventing the wheel?. Advanced materials, 16(14), 1151-1170.

5. Doradla, P., & Giles, R. H. (2014, March). Dual-frequency characterization of bending loss in hollow flexible terahertz waveguides. In SPIE OPTO (pp. 898518-898518). International Society for Optics and Photonics.
6. Doradla, P., Alavi, K., Joseph, C. S., & Giles, R. H. (2015, April). Flexible waveguide enabled single-channel terahertz endoscopic system. In SPIE OPTO (pp. 93620D-93620D). International Society for Optics and Photonics.
7. Kumar, A., Doradla, P., Narkhede, M., Li, L., Samuelson, L. A., Giles, R. H., & Kumar, J. (2014). A simple method for fabricating silver nanotubes. RSC Advances, 4(69), 36671-36674.
8. Doradla, P., Alavi, K., Joseph, C. S., & Giles, R. H. (2014, March). Terahertz polarization imaging for colon cancer detection. In SPIE OPTO (pp. 89850K-89850K). International Society for Optics and Photonics.
9. Doradla, P., Alavi, K., Joseph, C., & Giles, R. (2013). Detection of colon cancer by continuous-wave terahertz polarization imaging technique. Journal of Biomedical Optics, 18(9), 090504-090504.
10. Doradla, P., Alavi, K., Joseph, C. S., & Giles, R. H. (2013, March). Continuous wave terahertz reflection imaging of human colorectal tissue. In SPIE OPTO (pp. 86240O-86240O). International Society for Optics and Photonics.
11. Stiles, P. L., Dieringer, J. A., Shah, N. C., & Van Duyne, R. P. (2008). Surface-enhanced Raman spectroscopy. Annu. Rev. Anal. Chem., 1, 601-626.
12. Doradla, P., Alavi, K., Joseph, C. S., & Giles, R. H. (2016, May). Development of terahertz endoscopic system for cancer detection. In SPIE OPTO (pp. 97470F-97470F). International Society for Optics and Photonics.
13. Doradla, P. (2014). Development of single channel terahetz endoscopic system for cancer detection (Doctoral dissertation, University of Massachusetts Lowell).
14. Doradla, P., Alavi, K., Joseph, C., & Giles, R. (2014). Single-channel prototype terahertz endoscopic system. Journal of biomedical optics, 19(8), 080501-080501.